\documentclass[aps,twocolumn,prl,superscriptaddress,preprintnumbers,showpacs,tightenlines]{revtex4}
\usepackage{amssymb}
\usepackage{epsfig,graphicx,times}
\usepackage{amsmath}
\usepackage{mathrsfs}

\begin{document}

\title{Structural instabilities and mechanical properties of U$_2$Mo from first principles calculations}

\author{Ben-Qiong Liu}
\email{losenq@caep.cn}
\affiliation{Key Laboratory of Neutron Physics, Institute of Nuclear Physics and Chemistry, CAEP, Mianyang 621900, Sichuan Province, P.R. China}

\author{Xiao-Xi Duan}
\affiliation{Research Center of Laser Fusion, CAEP, Mianyang 621900, P.R. China}

\author{Guang-Ai Sun}
\affiliation{Key Laboratory of Neutron Physics, Institute of Nuclear Physics and Chemistry, CAEP, Mianyang 621900, Sichuan Province, P.R. China}

\author{Jin-Wen Yang}
\affiliation{Institute of Atomic and Molecular Physics, Sichuan University, Chengdu 610065, P.R. China}

\author{Tao Gao}
\affiliation{Institute of Atomic and Molecular Physics, Sichuan University, Chengdu 610065, P.R. China}


\begin{abstract}
We perform detailed first principles calculations of the structural parameters at zero pressure and high pressure, the elastic properties, phonon dispersion relation, and ideal strengths of U$_2$Mo with $C11_b$ structure. In contrast to previous theoretical studies, we show that the $I4/mmm$ structure is indeed a mechanically and dynamically unstable phase, which is confirmed by the negative elastic constant $C_{66}$ as well as the imaginary phonon modes observed along the $\Sigma_1$-N-P line. The calculations of ideal strengths for U$_2$Mo are performed along [100], [001], and [110] directions for tension and on (001)[010] and (010)[100] slip systems for shear load. The ideal shear strength is about 8.1 GPa, much smaller than tension of 18-28 GPa, which indicates that the ductile U$_2$Mo alloy will fail by shear rather than by tension.
\end{abstract}

\maketitle

\section{I. INTRODUCTION}

Uranium has received a lot of attention for its unique nuclear properties and its various applications in nuclear industry. At high temperature (1049 K$<T<$1405 K), the crystal structure of solid uranium is body-centered-cubic (bcc, $\gamma$ phase). On cooling, uranium experiences a solid-solid phase transformation to the body-centered tetragonal (bct) structure ($\beta$ phase) and then to the complex orthorhombic structure ($\alpha$ phase, $Cmcm$, 4 atoms per unit cell). The room temperature $\alpha$-U has some shortcomings such as poor oxidation and corrosion resistance, low hardness and yield strength. In order to improve mechanical properties and corrosion resistance of uranium at room temperature while maintaining the high density, uranium is frequently alloyed with other elemental metals. Several elements such as Mo \cite{B.W.Howlett}, Zr \cite{Y.Takahashi}, Nb \cite{B.-Q.Liu}, and Ti \cite{E.Kahana} exhibit a high degree of solid solubility in the high temperature $\gamma$ phase, therefore a wide range of metastable alloys can be formed at lower temperature \cite{C.N.Tupper}.

Mo exhibits a high solubility ($\sim$35 at.\%) in $\gamma$-U. Compared with other high density uranium alloys and compounds, the low-enriched uranium alloys with 6-12 wt.\% of Mo have attracted a great deal of attention and are recognized as the most prominent candidates for advanced research and test reactors, because they have a relatively larger $\gamma$ phase region and present more stable irradiation performance \cite{J.-M.Park,D.E.Burkes}. For trans-uranium-burning advanced fast nuclear reactors, it is shown that Mo is preferable to Zr because it is a stronger $\gamma$-stabilizer which provides stable swelling behavior in U-Pu-Mo fuels \cite{A.Landa2011}. And from a safety point of view, U-Pu-Mo fuels win an advantage over U-Pu-Zr fuels due to the higher thermal conductivity, lower thermal expansion, and higher melting points.

Although there have been extensive experimental studies on U-Mo alloys \cite{S.C.Parida,A.Leenaers,S.Van,V.P.Sinha}, theoretical attempts are rarely included. Alonso and Rubiolo \cite{P.R.Alonso} first evaluated the thermodynamic functions of U-Mo systems by employing first principles calculations with a cluster expansion technique, and predicted only one ground state compound, i.e., the bct U$_2$Mo with the $C11_b$ (MoSi$_2$ prototype) structure as its existence was observed experimentally. They also concluded that the stability of the $\gamma$U(Mo) phase was dominated by a three-body multisite interaction consisting of two pairs of first neighbours and one pair of third. Later, A. Landa \textit{et al.} \cite{A.Landa2011} studied the ground-state properties of U-Mo solid solutions by density functional theory. They revealed that there was a significant drop of the density of states in the vicinity of the Fermi level ($E_F$) of $C11_b$ U$_2$Mo which led to a decrease of the band-structure contribution to the total energy. It was suggested that the specific behavior could promote the stabilization of the U$_2$Mo compound. Whereafter, this work is extended to investigate ground-state properties of the bcc-based ($\gamma$) X-Mo (X=Np, Pu, and Am) solid solutions \cite{A.Landa2013}. The authors explained the reason for an increase of the heat of formation along the actinide row U-Mo$\rightarrow$Np-Mo$\rightarrow$Pu-Mo$\rightarrow$Am-Mo alloys, as well as the influence of magnetism on the deviation from Vegard's law for the equilibrium atomic volume. Recently, a density-functional theory study of mechanical and thermal properties of U$_2$Mo intermetallic has been reported \cite{S.Jaroszewicz}. The calculated elastic constants satisfied the mechanical stability criteria, implying the structural stability of $C11b$ U$_2$Mo. In this work, we present a comprehensive first principles study of the structural, elastic, lattice dynamical properties and ideal strength of the $C11_b$ U$_2$Mo. On the contrary, our results demonstrate that the structure is in fact unstable, which is in agreement with a very recent study \cite{X.Wang}.

This paper is arranged as follows. Section II describes details of the computational method. Section III is devoted to the calculations and discussions of structural, elastic properties, and phonon dispersion relation of U$_2$Mo, indicating the instability of this bct phase. In Section IV, the stress-strain relationships under tensile and shear loads are calculated. We conclude in Section V.

\section{II. Computational method}
First principles calculations are carried out by using the Vienna \textit{ab initio} simulation package (VASP) \cite{J.Hafner,G.Kresse,G.Kresser96}, with the frozen-core projector augmented wave (PAW) method. We use GGA descriptions for the exchange-correlation functional and set the cutoff energy of 650 eV in plane-wave basis expansion. Our plane wave cutoff energy is set to be much higher than 400 eV in Ref.\cite{X.Wang}, because in tensile tests the cell-volume change may cause discontinuous changes in stress values, if the cutoff energy is not enough. The $k$-point meshes in the Brillouin zone are sampled by $13\times13\times13$, determined according to the Monkhorst-Pack scheme. We use tetrahedron method with Bl\"{o}chl corrections for the energy calculation and Methfessel-Paxton's Fermi-level smearing to accelerate electronic structure relaxation, respectively. The quasi-Newton algorithm is adopted for the geometric relaxations, with a convergence criterion of the Hellmann-Feynman force being 0.01 eV/{\AA}.

To determine the formation energy of U$_2$Mo, we have also calculated the total energies of bcc U and Mo metals with the cutoff energy of 650 eV and 600 eV, the $k$-point meshes of 18$\times$18$\times$18 and 17$\times$17$\times$17, respectively. For accuracy, the $k$-mesh sampling has been increased to 17$\times$17$\times$17 during the elastic properties calculations. In the case of ideal shear strength calculation, a 20$\times$20$\times$20 $k$-mesh sampling has been applied due to the reason that shearing reduces the symmetry of the crystal and changes the shape of Brillouin Zone (BZ). There could be spurious changes in the energy if the $k$-point grid is too coarse.

\section{III. Structural instabilities}
\subsection{A. Structural properties}
The bct U$_2$Mo has a $D_{4h}^{17}$ ($I4/mmm$) space group (No. 139), with lattice parameters $a_0$=3.427 {\AA}, $c_0$=9.854 {\AA} giving a $c/a$ ratio of 2.876 \cite{A.E.Dwight}. When doing geometry optimization, we start with the experimental geometries and calculate the dependence of total energy ($E$) on the volume ($V$) of the phase so as to determine the bulk modulus $B_0$ and the equilibrium volume $V_0$. We calculate total energies for more than ten different volumes, and do a least-squares fit of the $E$-$V$ curves (shown in Fig.1) to the third-order Birch-Murnaghan equation of state \cite{F.Birch},
\begin{multline}
E(V)=-\frac{9}{16}B_0\Big[(4-B_0')\frac{V_0^3}{V^2}-(14-3B_0')\frac{V_0^{7/3}}{V^{4/3}}\\
+(16-3B_0')\frac{V_0^{5/3}}{V^{2/3}}\Big]+E_0,
\end{multline}
with $E_0$ being the equilibrium energy and $B_0'$ the pressure derivative of $B_0$.

The calculated lattice parameters are $a$=3.433 \AA, $c$=9.713 \AA, which are consistent with the experimental data and the very recent theoretical calculations \cite{X.Wang} (shown in Table I). The bulk modulus is obtained as $B_0=172.9$ GPa, which is in excellent agreement with Wang \emph{et al.}'s study \cite{X.Wang}, where $B_0$ is 182 GPa.

\begin{table}[htbp]
\begin{tabular}{p{55pt}p{30pt}p{30pt}p{30pt}p{30pt}p{40pt}}
\hline\hline
                                         &$a$({\AA}) &$c$({\AA}) &c/a       & $V_0$({\AA}$^3$) &$\Delta E_{form}$ (kJ/mol)\\
\hline
Expt. \cite{E.K.Halteman}                &3.427       &9.834     &2.8696    &115.49           &\\
FPLAPW \cite{P.R.Alonso}                 &3.44        &9.9       &2.8779    &117.15           &$\sim$ -6.13\\
FPLAPW \cite{S.Jaroszewicz}              &3.440       &9.631     &2.80      &113.97           &\\
EMTO \cite{A.Landa2011}                  &            &          &          &117.96           &$\sim$ -3.32\\
DFT \cite{X.Wang}                        &3.417       &9.714     &2.843     &113.42           &\\
Present work                             &3.433       &9.713     &2.829     &114.48           &-6.467\\
\hline\hline\\
\end{tabular}
\caption{Experimental and theoretical structural parameters for U$_2$Mo.}
\end{table}

\begin{figure}[tbp]
\includegraphics[width=8.0cm]{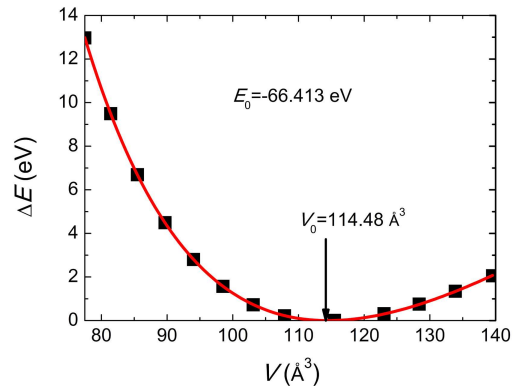}
\caption{(Color online) Total energy of U$_2$Mo as a function of unit-cell volume. The total energy at equilibrium is chosen as the zero of energy. The solid curve is a least-squares fit to the third-order Brich-Murnaghan equation of state \cite{F.Birch}. \label{FIG_1}}
\end{figure}

The formation energy of U$_2$Mo compound relative to the bcc U and Mo metals is also calculated as
\begin{equation}
\Delta E_{form}=E_{U_2Mo}-(E_U+E_{Mo}),
\end{equation}
where $E_{U_2Mo}$, $E_U$ and $E_{Mo}$ are the total energy of the compound, uranium and molybdenum, respectively. Our calculated formation energy of -6.467 kJ/mol is in agreement with the earlier theoretical result of -6.13 kJ/mol \cite{P.R.Alonso} but much lower than -3.32 kJ/mol \cite{A.Landa2011}.

Then we further study the crystallographic structure under compression. Figure 2 displays the calculated lattice parameters $a$ and $c$, as well as $V/V_0$ as functions of applied pressure at 0 K, respectively. One can see that the lattice parameter $c$ decreases faster with applied pressure than $a$ in Figure 2.

\begin{figure}[tbp]
\includegraphics[width=8.0cm]{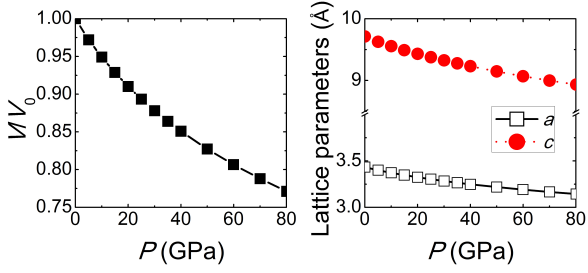}
\caption{(Color online) The lattice parameters $a$, $c$ and $V/V_0$ as a function of pressure. \label{FIG_2}}
\end{figure}

\subsection{B. Elastic properties}
After calculating the high-pressure structural properties of $C11_b$ U$_2$Mo, we further investigate the single-crystal elastic constants which determine the stiffness of a crystal against an externally applied strain. When the undistorted structure is deformed into a strain state $\mathbf{\varepsilon}$, the deformed lattice vectors can be obtained as $\mathbf{R'}=\mathbf{R}(\mathbf{I}+\mathbf{\varepsilon})$, where $\mathbf{I}$ is the identity matrix, and the strain tensor $\mathbf{\varepsilon}$ is defined as (Voigt's notation)
\begin{displaymath}
\mathbf{\varepsilon}=
\begin{pmatrix}
e_1    &e_6/2  &e_5/2\\
e_6/2  &e_2    &e_4/2\\
e_5/2  &e_4/2  &e_3\\
\end{pmatrix},
\end{displaymath}
where $e_i$ are the components of the strain vector $\mathbf{e}=(e_1,e_2,e_3,e_4,e_5,e_6)$. For small deformations, the total energy $E$ subjected to the strain can be expanded as
\begin{equation}
E=E_0+V_0\sum_{i=1}^6{\sigma_ie_i}+\frac{1}{2}V_0\sum_{i,j=1}^6{C_{ij}e_ie_j}+\cdots,
\end{equation}
with the stress vector $\mathbf{\sigma}=(\sigma_1,\sigma_2,\sigma_3,\sigma_4,\sigma_5,\sigma_6)$. For a tetragonal crystal there are only six independent elastic constants: $C_{11}=C_{22}$, $C_{12}$, $C_{13}=C_{23}$, $C_{33}$, $C_{44}=C_{55}$, and $C_{66}$. The others are either zero or satisfy the general condition $C_{ij}=C_{ji}$. Individual elastic constants $C_{ij}$ can be determined by computing the total energy as function of specific strain states. For instance, $\mathbf{e}(\delta)=(0,0,0,0,0,\delta)$ corresponds to the shear deformation along the 6-direction, and for a bct crystal Eq.(3) is reduced to $E(\delta)=E_0+\frac{V_0}{2}C_{66}\delta^2$, which allows a direct calculation of $C_{66}$ by fitting the $E$-$\delta$ curve. We list the different strain configurations to obtain the six independent elastic constants of bct crystal systems in Table II. For each case listed in Table II, we have used 13 different values of $\delta$: $\delta$=0, $\pm0.003$, $\pm0.006$, $\cdots$, $\pm0.018$.
\begin{table}[htbp]
\begin{tabular}{p{20pt}p{90pt}p{110pt}}
\hline\hline
       &Strain configuration                                  &$\Delta E/V_0$ to $O(\delta^2)$\\
\hline
1      &$\mathbf{e}=(\delta,\delta,0,0,0,0)$                  &$(C_{11}+C_{12})\delta^2$\\
2      &$\mathbf{e}=(0,0,0,0,0,\delta)$                       &$\frac{1}{2}C_{66}\delta^2$\\
3      &$\mathbf{e}=(0,0,\delta,0,0,0)$                       &$\frac{1}{2}C_{33}\delta^2$\\
4      &$\mathbf{e}=(0,0,0,\delta,\delta,0)$                  &$C_{44}\delta^2$\\
5      &$\mathbf{e}=(\delta,\delta,\delta,0,0,0)$             &$(C_{11}+C_{12}+2C_{13}+\frac{C_{33}}{2})\delta^2$\\
6      &$\mathbf{e}=(0,\delta,\delta,0,0,0)$                  &$(\frac{C_{11}}{2}+C_{13}+\frac{C_{33}}{2})\delta^2$\\
\hline\hline\\
\end{tabular}
\caption{Parametrizations of strains used to calculate the six independent elastic constants of bct U$_2$Mo.}
\end{table}

In Table III, we summarize our theoretical calculated elastic constants, and compare with previous theoretical results. It is shown that our calculated results are in reasonable agreement with the very recent calculation \cite{X.Wang}. It is worth noting that we have obtained a negative $C_{66}$ of -34.6 GPa, whereas Jaroszewicz \emph{et al.} \cite{S.Jaroszewicz} gave the value of $C_{66}$ of $\sim$20 GPa and demonstrated mechanical stability of this structure.

\begin{table}[htbp]
\begin{tabular}{p{70pt}p{20pt}p{20pt}p{20pt}p{20pt}p{20pt}p{20pt}}
\hline\hline
                               &$C_{11}$  &$C_{12}$  &$C_{13}$  &$C_{33}$  &$C_{44}$  &$C_{66}$\\
\hline
DFT\cite{X.Wang}               &255       &165       &124       &302       &37        &-12\\
FPLAPW \cite{S.Jaroszewicz}    &254       &161       &125       &295       &38        &$\sim20$\\
Present work                   &243.7     &148.9     &148.5     &285.1     &17.2      &-34.6\\
\hline\hline\\
\end{tabular}
\caption{Calculated elastic constants (in GPa) of bct U$_2$Mo.}
\end{table}

In order to examine the influence of these elastic constants $C_{ij}$s on the mechanical stability, we refer to the well-known Born-Huang stability criteria \cite{M.Born} for the tetragonal crystal systems:
\begin{multline}
(2C_{11}+C_{33}+2C_{12}+4C_{13})>0,\\
C_{11}>0, C_{33}>0, C_{44}>0, C_{66}>0,\\
(C_{11}-C_{12})>0, (C_{11}+C_{33}-2C_{13})>0.
\end{multline}
From Table III we can see that all the predicted elastic constants satisfy these conditions except $C_{66}<0$, indicating that the tetragonal phase is mechanically unstable at zero temperature, which is consistent with Wang \emph{et al.}'s results \cite{X.Wang} where $C_{66}=-12$ GPa. We note that in Jaroszewicz \emph{et al.}'s study \cite{S.Jaroszewicz}, the calculated value of $C_{66}$ is very small and suggests that the tetragonal phase is in fact unstable against the shear deformation along the direction of the $C_{66}$ elastic tensor.

Based on the obtained six single-crystal elastic constants of U$_2$Mo, one can further study its noteworthy polycrystalline elastic properties according to Voigt-Reuss-Hill approximation \cite{D.W.Voigt,A.Reuss,R.Hill}. In Voigt theory, the bulk ($B_V$) and shear modulus ($G_V$) can be obtained as
\begin{equation}
B_V=\frac{2C_{11}+C_{33}+2C_{12}+4C_{13}}{9},
\end{equation}
\begin{equation}
G_V=\frac{2C_{11}+C_{33}-C_{12}-2C_{13}+6C_{44}+3C_{66}}{15},
\end{equation}
while in Reuss theory,
\begin{equation}
\frac{1}{B_R}=2(S_{12}+S_{23}+S_{13})+S_{11}+S_{22}+S_{33},
\end{equation}
\begin{multline}
\frac{1}{G_R}=\frac{1}{5}(S_{44}+S_{55}+S_{66})-\frac{4}{15}(S_{12}+S_{23}+S_{13})\\
+\frac{4}{15}(S_{11}+S_{22}+S_{33}),
\end{multline}
where $S_{ij}$ are the elastic compliance constants, derived from the inverse of $C_{ij}$s. Then the elastic modulus of the polycrystalline aggregates can be given by Hill's average $B=\frac{1}{2}(B_V+B_R)$ and $G=\frac{1}{2}(G_V+G_R)$. The Young's modulus $Y$, and the Poisson's ration $\nu$ can be calculated as
\begin{equation}
Y=\frac{9BG}{3B+G}, \nu=\frac{3B-2G}{2(3B+G)}.
\end{equation}

Employing the relations above, the calculated bulk modulus, shear modulus, Young's modulus and Poisson's ratio for U$_2$Mo are summarized in Table IV. For polycrystalline phases, Pugh \cite{S.F.Pugh} has introduced the ratio of bulk to shear modulus $B/G$ by considering that the shear modulus $G$ represents the resistance to plastic deformation, while the bulk modulus $B$ represents the resistance to fracture. A high (low) $B/G$ value is associated with ductility (brittleness). Although this parameter is generally applied for cubic materials, it is interesting to examine U$_2$Mo. The $B/G$ value for the brittle bcc-U is 0.66, whereas for bcc-Mo is 1.09. In present work the value of U$_2$Mo is as high as 5.9, suggesting high ductility of this compound.

\begin{table}[htbp]
\begin{tabular}{p{30pt}p{50pt}p{32pt}p{32pt}p{20pt}p{32pt}p{20pt}}
\hline\hline
                                &                                   &$B$ (GPa)           &$G$ (GPa)    &$B/G$  &$Y$ (GPa) &$\nu$\\
\hline
bcc-U                           &Expt.\cite{D.Roundy}               &134                 &203.01       &0.66   &          &\\
                                &Present work                       &132.2               &204.4        &0.647  &          &\\
bcc-Mo                          &Expt.\cite{G.Simmons,J.Donohue}    &267.67              &245.72       &1.09   &          &\\
                                &Present work                       &261.4               &234.3        &1.12   &          &\\
U$_2$Mo                         &FPLAPW method \cite{S.Jaroszewicz} &180                 &36           &5      &102       &0.4\\
U$_2$Mo                         &Present work                       &184.4               &21.7         &5.9    &62.64     &0.44\\
\hline\hline\\
\end{tabular}
\caption{The calculated bulk modulus $B$, shear modulus $G$, $B/G$, Young's modulus $Y$ and Poisson's ratio $\nu$ of bcc-U, bcc-Mo, U$_2$Mo.}
\end{table}

\subsection{C. Phonon dispersion relation}
As stability requires that the energies of phonons be positive for all the wave vectors in the Brillouin Zone \cite{D.M.Clatterbuck}, the dynamical instability of U$_2$Mo can be investigated by the full phonon dispersion relation. The calculations are performed with the code of Phonopy \cite{A.Togo2008,A.Togo2010} by constructing a $2\times2\times2$ supercell. The full phonon dispersion of U$_2$Mo consists of 18 branches. Results of our calculations along the high-symmetry lines of the bct Brillouin zone are shown in Fig.3. The imaginary frequencies of the unstable modes are represented as negative values. Lattice instabilities are observed to occur around the high-symmetry directions $\Sigma_1$-N-P, where the phonon branches exhibit highly large imaginary frequencies, suggesting that this structure is dynamically unstable.

\begin{figure}[tbp]
\includegraphics[width=8.0cm]{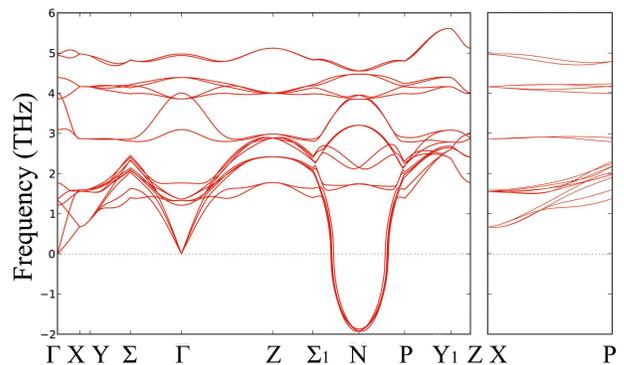}
\caption{(Color online) Phonon dispersion for U$_2$Mo along symmetry lines in the body centered tetragonal BZ. The imaginary frequencies of the unstable modes are plotted as negative values. \label{FIG_2}}
\end{figure}

\section{IV. Stress-strain relationship and ideal strength}
In view of the great interest of the study of the intrinsic (ideal) mechanical properties of materials, in this Section we focus on the intrinsic stress-strain relationship of U$_2$Mo under tension and shear. The term "intrinsic" refers to bulk perfect crystal without any defects. The ideal strength, which provides an upper bound of stress that the material can attain under uniform deformation without any extrinsic effects, has been recognized as an essential mechanical parameter of material \cite{J.Wang,S.Ogata}. It has obtained considerable attention theoretically and experimentally \cite{J.Li}. This is first because it is sometimes useful to know the highest strength a particular material can possibly have. Secondly, this upper bound is actually reached or closely approached in a number of experiments of nano-structures or nano regions thanks to the development of material processing techniques. The most common case is deformation via stress-induced phase transformations, since some ductile metals and alloys could approach the limit of strength in the region of stress concentration before a crack \cite{J.W.M.Jr}. Moreover, with recent advances in modern computational techniques, first principles studies based on density-functional theory have been successfully employed to provide quantitatively believable estimates of these upper limits \cite{R.F.Zhang,J.Wang}.

For the determination of the theoretical strength of a perfect crystal, one has to calculate the stress-strain curves for large strains which yield the ideal strengths. We compute the stress-strain dependence by incrementally deforming the modeled cell in the applied strain direction. The atomic basis vectors perpendicular to the applied strain are simultaneously relaxed until the other components vanish. Meanwhile, all the internal freedoms of the atom are fully relaxed at each step. To ensure that the strain path is continuous, the starting atomic position at each strain step are taken from the relaxed coordinates of the previous smaller strain step. According to \cite{D.Roundy}, the uniaxial tensile stress $\sigma$ is derived from
\begin{equation}
\sigma=\frac{1+\varepsilon}{V(\varepsilon)}\frac{\partial E(\varepsilon)}{\partial\varepsilon},
\end{equation}
and the shear stress $\tau$ is given by
\begin{equation}
\tau=\frac{1}{V(\gamma)}\frac{\partial E(\gamma)}{\partial\gamma}
\end{equation}
where $E(\varepsilon)$ is the strain energy, $V(\varepsilon)$ and $V(\gamma)$ are the volumes at the corresponding tensile strain $\varepsilon$ and shear strain $\gamma$, respectively.

In order to check the reliability of our method, we have firstly calculated the ideal tensile strength of bcc Mo in the [100] direction (shown in Figure 4), since there has been some theoretical results to compare with \cite{W.Luo}. Figure 4 indicates that our results are in perfect agreement with Luo \textit{et al.}'s \cite{W.Luo} study and confirms our method.
\begin{figure}[tbp]
\includegraphics[width=8.0cm]{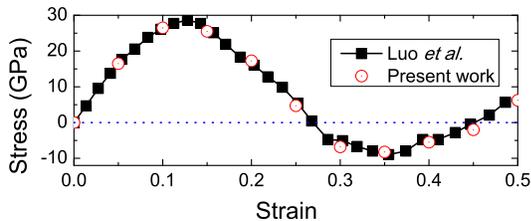}
\caption{(Color online) The stress-strain curve of Mo under [100] tension. The black squares represent Luo \textit{et al.}'s calculation \cite{W.Luo}. \label{FIG_X}}
\end{figure}

The calculations of ideal tensile strengths for U$_2$Mo are performed along [100], [001] and [110] crystallographic directions, respectively. Figure 5 shows the unit cell and the redefined supercell of bct U$_2$Mo adopted in our calculation. The unit cell (black frame) is used to calculate the [100] and [001] tensile stress-strain curves; while the supercell (red frame) with specific crystal directions is adopted to simulate the tension in [110] direction. Much finer meshes of $17\times17\times17$ are used to sample the Brillouin Zone when carrying out cell and atomic relaxations.

\begin{figure}[tbp]
\includegraphics[width=8.0cm]{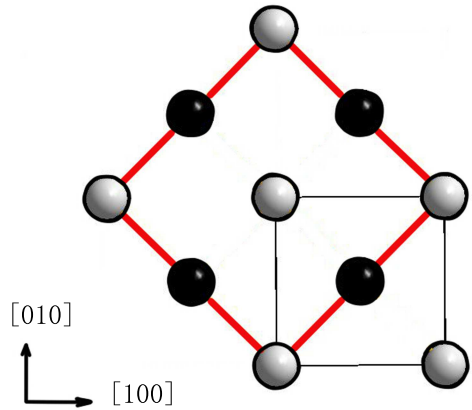}
\caption{(Color online) A schematic projection of U$_2$Mo in the \textit{ab} plane, with Mo shown in grey, U in black. The black profile is the unit cell with six atoms. The red one is the redefined orthorhombic supercell, with [110], [$\bar{1}$10], and [001] representing the three orthogonal lattice vectors $a'$, $b'$, and $c'$, respectively. \label{FIG_2}}
\end{figure}

Fig.6 shows the stress-strain and energy-strain relationships of U$_2$Mo loaded in tension, respectively. It can be seen that the ideal tensile strength are about 18-28 GPa. The tensile strength along the [001] direction is about 27.6 GPa, higher than those in [100] ($\sim18.1$ GPa) and [110] ($\sim20.5$ GPa) directions. Moreover, the shape of stress-strain curves in different directions are different from each other. The stress-strain curves are discontinuous except [110] tension. There is only one maximum in [110] tension, the subsequent stress smoothly decreases. However, for the other two cases, the stress-strain curves become much more complex that at first the stress is incrementally increasing until the energy reaches an inflection point, and subsequently exhibits more characteristics. For [100] tension, a rapid decrease occurs at a strain of 0.15; then the stress increases smoothly up to a second maximum at strain of about 0.3, after that it changes its sigh from positive to negative at strain between 0.38 and 0.4. This transition in stress have close relation with structural changes in deformed U$_2$Mo as shown in Fig.7.

\begin{figure}[tbp]
\includegraphics[width=8.9cm]{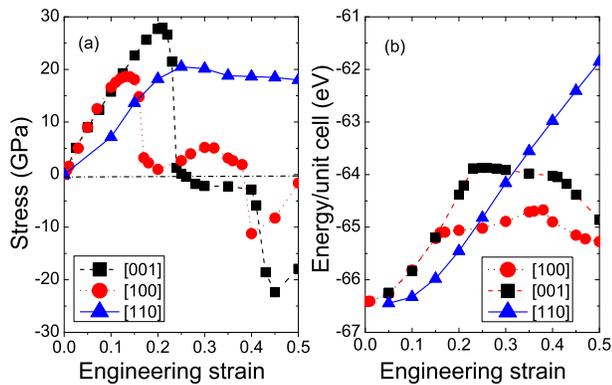}
\caption{(Color online) Stress-strain (a) and energy-strain (b) curves of U$_2$Mo under tensile load. \label{FIG_X}}
\end{figure}

We show the behavior of lattice vectors during [100] tension in Figure 7. The [100] lattice vector is always controlled to increase linearly during tension; whereas [010] and [001] lattice parameters reveal the decaying tendency. Nevertheless, an abrupt increase of the length of lattice vector $c$ but a decrease of $b$ appears from strain of 0.38 to 0.4, i.e., an abrupt increase of the ratio $c/b$ from 2.61 to 3.05 corresponds to the transition in stress from positive value to negative.

\begin{figure}[tbp]
\includegraphics[width=8.9cm]{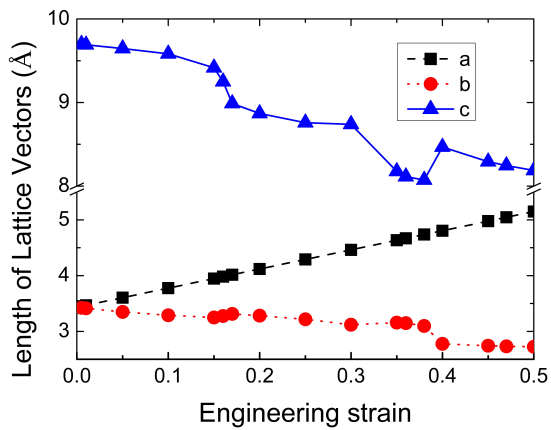}
\caption{(Color online) The length of lattice vectors as a function of tensile strain along [100] direction. An abrupt change occurs between strain of 0.38 and 0.4, corresponding to the transition of stress from positive to negative value as shown in Fig.6.\label{FIG_X}}
\end{figure}

For simple shear deformations, the volumes are fixed at the initial equilibrium volumes, whereas they are changed under shear load. For the slip modes of (001)[010], the calculated stress shows an oscillating and decaying behavior, and the ideal strength is 8.1 GPa with the critical strain of $\sim$0.3. Abrupt changes appear in the region [0.3, 0.45], which have close relation with structural changes in the deformed U$_2$Mo as shown in Fig.8(b). Compared with tension, the shear energy-strain relationship is consistently below that of tensile curves. It indicates that the shear modulus of U$_2$Mo in the considered slip system is smaller than Young's modulus. For the slip mode of (010)[100], the calculated stress is negative at the very beginning, which implies that the $C11_b$ structure is unstable under shear load on (010)[100] slip system and is consistent with the negative $C_{66}$.

\begin{figure}[tbp]
\includegraphics[width=8.9cm]{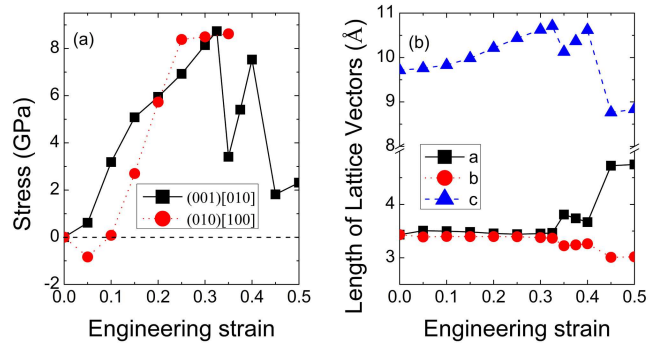}
\caption{(Color online) Stress-strain relationship (a) and the length of vectors as a function of strain on (001)[010] slip system (b) under shear load. \label{FIG_X}}
\end{figure}

\section{V. Conclusions}

In summary, we have comprehensively studied the structural, elastic properties, lattice dynamics and ideal strength of the experimentally observed $C11_b$ U$_2$Mo alloy by first principles calculations. In contrast to the previous theoretical studies, our results imply that U$_2$Mo is unstable since the elastic constant $C_{66}$ is negative and violates the well-known Born-Huang stability criteria. Another evidence is provided by phonon instabilities along $\Sigma_1$-N-P. This could be a typical problem that the high-temperature phase is mechanical unstable at low temperatures because on the basis of experimental observation, the $C11_b$ structure is stable at temperature of $\sim873$ K. Recently, P. S\"{o}derlind \emph{et al.} \cite{P.Soderlind} have developed self-consistent \emph{ab initio} lattice dynamics (SCAILD) scheme in conjunction with highly accurate and fully relativistic density functional theory, which could model high-temperature lattice dynamics and seems to be a possible solution to this long-standing difficulty. In spite of unstable at 0 K, the vibrational entropy introduced through SCAILD might produce a stable phonon dispersion at the experimental temperature. Further study is required in near future.

Moreover, we have studied the ideal strength of U$_2$Mo under tensile and shear loads, respectively. The ideal shear strength for (001)[010] is about 8.1 GPa, much smaller than tension of 18-28 GPa, implying that U$_2$Mo will fail by shear rather than by cleavage. This results is consistent with the above-mentioned that the $C11_b$ structure is unstable against the shear deformation along the direction of $C_{66}$ elastic tensor.

\section*{ACKNOWLEDGMENTS} This work was supported by the National Natural Science Foundation of China (Grant No. 11305150), and Science and Technology Foundation of China Academy of Engineering Physics (No.2013A0302013). 

\end{document}